\documentclass[oneside,twocolumn,aps,pra,preprintnumbers,bibnotes10pt,superscriptaddress,amsmath,amssymb]{revtex4-2}
\usepackage{graphicx,subfigure,wrapfig}
\usepackage{float}
\usepackage{color}
\usepackage[dvipsnames]{xcolor}
\usepackage{epstopdf}
\usepackage{amsmath}
\usepackage{braket}
\usepackage{amsthm}
\usepackage{amssymb}
\usepackage{amsfonts}
\usepackage{graphicx,subfigure}
\usepackage{txfonts}
\usepackage{ulem}
\usepackage{mathtools}
\usepackage{bm}
\usepackage{bbm}
\usepackage{hyperref}
\usepackage{relsize}
\usepackage[hang,flushmargin]{footmisc} 
\usepackage{comment}
\usepackage{cancel}

\newcommand{\remove}[1]{}                                                 

\definecolor{giu}{rgb}{1,0.7,0.7}

\newcommand{\be}{\begin{equation}}
\newcommand{\ee}{\end{equation}}
\newcommand{\beq}{\begin{eqnarray}}
\newcommand{\eeq}{\end{eqnarray}}

\begin{document}

\title{Increasing the stability of a superfluid in a rotating necklace potential}



\author{Giulio~Nesti}
\affiliation{European Laboratory for Nonlinear Spectroscopy (LENS), University of Florence, 50019 Sesto Fiorentino, Italy}

\author{Luca Pezzè}
\affiliation{European Laboratory for Nonlinear Spectroscopy (LENS), University of Florence, 50019 Sesto Fiorentino, Italy}
\affiliation{Istituto Nazionale di Ottica del Consiglio Nazionale delle Ricerche (CNR-INO), Largo Enrico Fermi 6, 50125 Firenze, Italy}

\begin{abstract}

Recent experiments have probed the stability of ring superfluids in the presence of Josephson barriers or Gaussian impurities. 
Here we present a theoretical analysis that extends beyond the regimes explored so far.
We study the onset of dynamical instabilities in a ring superfluid, addressing both tunneling and  hydrodynamic regimes.
The stability of the system is controlled by the effective rotation frequency $\omega$, given by the difference between the initial quantized circulation and the frequency of barrier rotation.
The instability occurs when $\omega$ overcomes a critical value $\omega_c$.
We show that $\omega_c$ increases approximately linearly with the number of barriers, with a slope set by the barrier height and width.
When the system is quenched into the dynamically unstable regime, it emits multiple solitons, which can switch or even reverse the direction of circulation. 
The stabilization mechanism is robust against imperfections of the potential and does not require a perfectly periodic array of barriers. 
In particular, we find that adding a disordered speckle potential to an ordered array of barriers can further increase $\omega_c$: disorder can therefore make a ring superfluid more resilient to dynamical instabilities.

\end{abstract}

\maketitle

\section{Introduction}

Degenerate quantum gases are ideal clean platforms to study several fascinating aspects associated with the interplay of superfluidity and impurity-induced dissipation~\cite{Landau1941, LeggettRMP2001, Pethick_Smith_2008, WuPRA2001, AstrakharchikPRA2004, SykesPRL2009, PinskerPA2017}.
%
Early demonstrations of superfluidity in these systems ~\cite{RamanPRL1999, MatthewsPRL1999, OnofrioPRL2000, CataliottiSCIENCE2001, KinastPRL2004, ZwierleinNATURE2005, DesbuquoisNATPHYS2012} have paved the way to recent quantification of the superfluid fraction~\cite{Legget1970, LeggPhys1998} in periodic potentials~\cite{Chauveau2023} and supersolids~\cite{BiagioniNATURE2024}.
Controlled dissipation can be induced by external optical lattices~\cite{MorschPRL2001, FallaniPRL2004}, disordered potentials~\cite{ LyePRL2005, ClémentPRL2005, PasienskiNATPHYS2010}, or Gaussian impurities~\cite{DesbuquoisNATPHYS2012, KwonPRA2015, KwonPRL2016}.

\begin{figure}[hb!]
\centering
\includegraphics[width=\columnwidth]{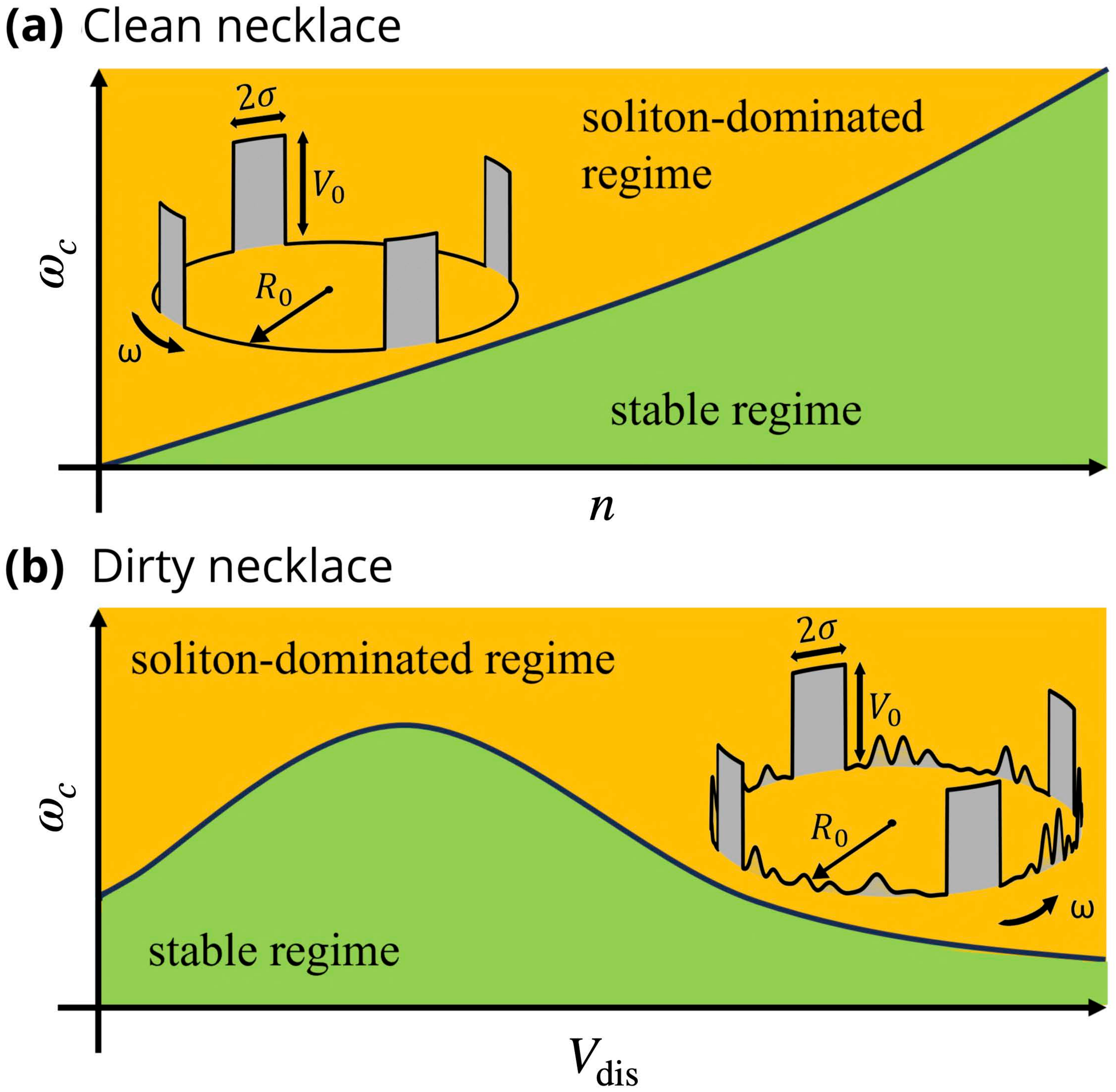}
\caption{Schematic summary of the main results of this work.
Our system is a 1D ring BEC including potential barriers rotating at an effective frequency $\omega$.
We distinguish a regime where the BEC supports stationary (stable) solutions (green region) and a regime where the system becomes dynamically unstable due to soliton emission (yellow region).
These are separated by a critical frequency $\omega_c$ (black line).
(a) Clean necklace.
The potential consists of $n$ identical barriers of height $V_0$ and half-width $\sigma$ (inset).
We find that $\omega_c(n)$ increases monotonically with $n$, with a slope set by the interaction strength and the height and the width of the barriers.
(b) Dirty necklace.
Here a disordered background potential of maximum amplitude $V_{\rm dis}$ is superposed to the clean necklace (inset).
For fixed $n$, we find that $\omega_c$ increases further with $V_{\rm dis}$, reaching a maximum for $V_{\rm dis} \approx V_0$. This schematic summarizes the generic behavior observed for an arbitrary number of barriers $n$.
}
\label{Fig1}
\end{figure}

A particularly interesting platform is provided by ultracold gases in ring traps~\cite{GuptaPRL2005,RyuPRL2007,TononiNRP2023}.
In the absence of additional external potentials, ring superfluids can support current states with finite quantized circulation. 
This has been experimentally demonstrated in both bosonic~\cite{RyuPRL2007, MoulderPRA2012, CormanPRL2014} and fermionic~\cite{CaiPRL2022,DelPacePRX2022} ultracold gases, with persistent currents of multiple circulation quanta and lifetimes of the order of seconds.
Ring traps can be further equipped with additional external potentials. 
Experiments with a single rotating barrier or weak link have investigated the onset of phase slip processes~\cite{WrightPRL2013}, hysteresis~\cite{EckelNATURE2014} and current–phase relation~\cite{EckelPhys2014}.
Superfluidity holds only up to a critical relative velocity between the fluid and the potential: above this threshold, excitations become energetically favorable and the flow becomes dissipative~\cite{PiazzaPRA2009, RamanathanPhys2011, WrightPRL2013, WrightPRA2013, PoloPRL2019, XhaniATOM2023}.
More recently, the experimental and theoretical attention has been devoted to the ring superfluids with several potential barriers. 
In Ref.~\cite{RyuPhys2013, RyuNature2020} a ring with two movable barriers have realized a superfluid analogue of SQUIDs.
Reference~\cite{PezzeNATCOMM2024} investigated the transport properties in a toroidal Bose-Einstein condensate (BEC) equipped with a controllable number, $n$, of tunneling barriers. 
It was shown that, in the tunneling regime ($V_{0}/\mu>1$), the Josephson critical current $J_{c}$ increases with $n$.
A similar effect has been observed in further experiments with a toroidal BEC and a variable number of localized Gaussian impurities~\cite{XhaniImpurities2025} of size comparable with the BEC healing length.

In this manuscript, we extend the characterization of necklace superfluids beyond the regime of parameters and phenomena studied so far~\cite{PezzeNATCOMM2024, XhaniImpurities2025}.
We consider a one-dimensional (1D) BEC with circulation $\nu$ in a ring equipped with $n$ barriers rotating at frequency $\Omega$.
We address two different scenarios, see Fig.\ref{Fig1}. 
In the so-called {\it clean necklace}, Fig.\ref{Fig1}(a), the barriers are identical and periodically arranged.
For $\omega$ below a critical value, $\omega_c$, the system exhibits a stable ground state flow. 
For $\omega > \omega_c$, a dynamical instability is triggered when the local phase jump -- or, equivalently, the maximum superfluid velocity -- exceeds a critical threshold.
A square-barrier model provides analytical insights to the instability mechanism and yields an explicit expression for $\omega_c$.
A key result is that $\omega_c$ increases with the number of barriers $n$, as schematically indicated in Fig.~\ref{Fig1}(a). 
This enhanced stability is a consequence of circulation quantization in the ring: the phase drop across each barrier scales as $\delta\phi \sim 1/n$, thereby suppressing the onset of phase slips as $n$ increases. 
Although a similar topological mechanism was previously identified in the scaling of the maximum sustainable current $J_c$ in the clean necklace~\cite{PezzeNATCOMM2024}, we emphasize that $J_c$ and $\omega_c$ characterize distinct properties of the system.
In particular, the maximum current does not generally occur at $\omega = \omega_c$.
We show that the increase of $\omega_c$ with $n$ is robust over a wide range of interaction strengths and barrier parameters (height $V_0$ and width $\sigma$).
This behavior persists in the hydrodynamic regime ($\xi \ll \sigma$, with $\xi$ being the healing length), in the Josephson regime ($V_0/\mu > 1$, with $\mu$ being the chemical potential), and in the weak-link regime ($V_0/\mu \le 1$, $\xi \ge \sigma$), while being insensitive to the detailed shape of the barriers. 
For quenches to $\omega > \omega_c$, we observe a soliton-dominated response in which $n$ solitons -- one per barrier -- are generated simultaneously. 
By varying $n$, the resulting dynamics can reverse the initial circulation, enabling the realization of a superfluid switch or inverter.
We also analyze a {\it dirty necklace} configuration, Fig.~\ref{Fig1}(b), in which a disordered potential is present between the barriers. 
We find that $\omega_c$ further increases with the height of the disordered potential, $V_{\rm dis}$, relative to the barriers height.
A resonant response occurs for $V_{\rm dis}\approx V_0$ where the $\omega_c$ reaches a maximum.  
Thus, contrary to common-sense expectations, weak disorder enhances the stability of the necklace superfluid against dynamical instabilities: this enhancement is not due to the randomness itself but from adding more barriers.
As in the clean case, quenches into the unstable regime allow for controlled inversion of the superfluid circulation and tailored vortex emission.

\section{Superfluid Necklace}

The BEC is confined to a 1D ring of radius $R$ and equipped with a repulsive potential $V(\theta)$, where $\theta$ is the azimuthal angle. 
Following Ref.~\cite{PezzeNATCOMM2024}, we refer to this configuration as a {\it superfluid necklace}. 
The explicit form of $V(\theta)$, specified below, distinguishes between clean and disordered (“dirty”) realizations. 
We investigate the stability of the BEC when the potential is rotated at angular velocity $\Omega$ with respect to the laboratory frame.
In the co-rotating frame, the dimensionless Gross-Pitaevskii equation (GPE) reads
\begin{equation}
\resizebox{0.9\linewidth}{!}{$
\biggl[
    -\dfrac{1}{2}\dfrac{\partial^{2}}{\partial \theta^{2}}
    + g|\Psi(\theta,t)|^2
    + V(\theta)
    + i \Omega\, \dfrac{\partial}{\partial \theta}
\biggr]\Psi(\theta,t)
= i \dfrac{\partial}{\partial t}\Psi(\theta,t)$}
\label{1DGPEFULL}
\end{equation}
where $\Psi(\theta,t)$ is the order parameter, normalized as
$\int_{0}^{2\pi} |\Psi(\theta,t)|^2 d\theta = 1$, and $g $ is the coupling constant.
In the following, we focus on the case of repulsive interactions, $g > 0$, and briefly discuss the case of attractive interaction, $g < 0$, in Appendix~\ref{Att}.
Energies, angular velocities and times are measured in units of $E_R \equiv \hbar^2/mR^2$, $\Omega_R \equiv \hbar/mR^2$ and $t_R \equiv mR^2/\hbar$, respectively, where $m$ is the atomic mass.
Due to the multiply connected geometry, the single-valuedness condition on the wavefunction imposes the quantization condition
\begin{equation}
\label{Quant}
\int_0^{2\pi} \upsilon(\theta,t)\, d\theta = 2\pi \nu(t),
\end{equation}
where
$\upsilon(\theta,t) = \partial_\theta \phi(\theta,t)$ is the superfluid angular velocity in units of $\upsilon_R = R \Omega_R = mR/\hbar$, $\phi(\theta,t)$ is the condensate phase and $\nu(t)$ is an integer winding number, which becomes time-independent for stable states (i.e. no soliton is observed in the dynamics, see below). 
\begin{figure}[t!]
\centering
\includegraphics[width=\columnwidth]{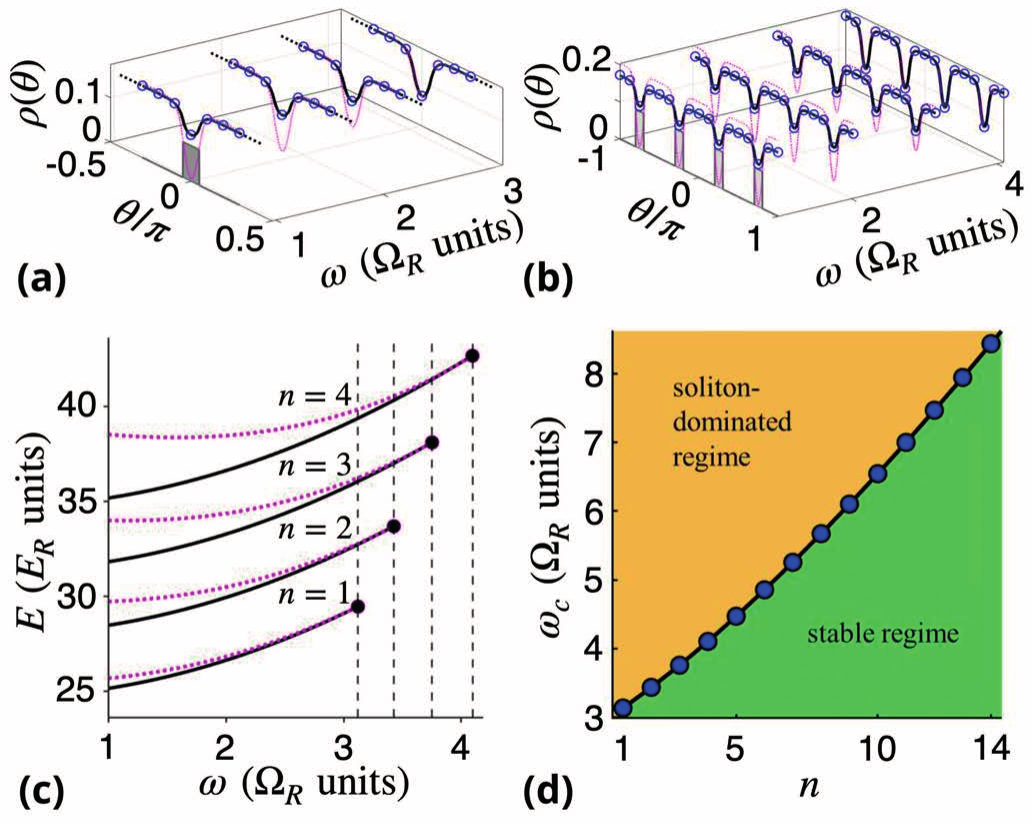}
\caption{
Top panels show the analytical density profiles of the ground (solid black) and excited (dotted pink) states, Eqs.~(\ref{solutionsins}) and~(\ref{solutionsout}), for $n = 1$ (a) and $n = 4$ (b), respectively. 
The density is plotted as a function of both the azimuthal angle $\theta$ and the effective rotation frequency $\omega$. 
Blue dots indicate the numerical ground-state density obtained by solving Eq.~(\ref{1DGPERes}). 
The gray shaded areas represent the external potential of Eq.~(\ref{Barriers}).
Panel (c) displays the mean-field energy, Eq.~(\ref{GPEEN}), of the ground (solid black) and excited (dotted pink) states as a function of $\omega$. 
Different curves correspond to different values of $n$ and are vertically offset for clarity. 
Vertical lines and filled dots highlight the critical frequency $\omega_c$.
Panel (d) shows $\omega_c(n)$ as a function of the number of barriers $n$ (blue dots); the solid line is a guide to the eye. Green (yellow) shading indicates the stable (unstable) region, as in Fig.~\ref{Fig1}(a). 
In all panels, the parameters are $g = 300$, $V_{0}/\mu \approx 0.4$, and $\sigma/\xi \approx 0.5$.
}
\label{Fig2}
\end{figure}
At equilibrium, the order parameter evolves in time through the chemical potential $\mu$,
\begin{equation}
\Psi(\theta,t) = \sqrt{\rho(\theta)}\, e^{i[\phi(\theta) - \mu t]},
\end{equation}
where $\rho(\theta)$ is the stationary density profile. 
With this ansatz, Eq.~(\ref{1DGPEFULL}) separates into two equations~\cite{Pethick_Smith_2008}. 
The imaginary part yields a continuity equation, which defines the superfluid current
\begin{equation}
\label{J}
J = \rho(\theta)\,[\upsilon(\theta) - \Omega]
\end{equation}
in the chosen dimensionless units. In the static case the current is spatially uniform along the ring, so that $\partial J/\partial \theta = 0$. 
Substituting Eq.~(\ref{J}) into Eq.~(\ref{Quant}) gives
\begin{equation}
\label{J2}
J = \frac{\omega f_s}{2\pi},
\end{equation}
where
\begin{equation}
\label{fs}
f_s = (2\pi)^2\biggl(\int_0^{2\pi} \frac{d\theta}{\rho(\theta)}\biggr)^{-1}
\end{equation}
is Leggett's superfluid fraction~\cite{Legget1970,LeggPhys1998,PezzeNATCOMM2024}, and where:
\begin{equation} \label{Eq.omega}
\omega = \nu - \Omega 
\end{equation}
is the effective rotation frequency. Since we are studying a stationary configuration, in Eq.~(\ref{Eq.omega}) we drop the time-dependence of $\nu$.
The real part of Eq.~(\ref{1DGPEFULL}) provides an  equation for the density,
\begin{equation}
\label{1DGPERes}
\mu_{0} \sqrt{\rho(\theta)} =
\biggl(
    -\dfrac{1}{2}\dfrac{\partial^{2}}{\partial \theta^{2}}
    + \dfrac{1}{2}\dfrac{J^{2}}{\rho(\theta)^2}
    + V(\theta)
    + g \rho(\theta)
\biggr)\sqrt{\rho(\theta)},
\end{equation}
where $\mu_{0} = \mu + \Omega^2/2$. 
Thus, the effect of rotation is fully absorbed into the current $J$, together with a shift of the chemical potential by $\Omega^2/2$, making Eq.~(\ref{1DGPERes}) formally equivalent to the stationary GPE of a non-rotating condensate.
The ground state of the system is obtained by solving Eqs.~(\ref{J}) and~(\ref{1DGPERes}) self-consistently, via imaginary-time propagation, starting from a uniform initial wavefunction.

\section{Clean necklace}

We define the clean superfluid necklace as the case of $n$ identical barriers uniformly distributed along the ring.
Throughout most of this work we consider square barriers, which allow for analytical solutions of the GPE (see below). 
The potential is then given by
\begin{equation}\label{Barriers}
    V(\theta) = 
    \begin{cases}
      V_{0}, & \text{for } \theta \in [\theta_j - \sigma, \theta_j + \sigma],\\
      0,     & \text{otherwise},
    \end{cases}
\end{equation}
for $j = 1,\dots,n$, where $n$ is the number of barriers, $V_{0}$ and $\sigma$ are the height and half-width of each barrier, respectively, and $\theta_j = 2\pi j/n$ is the position of the $j$-th barrier. 
An example of the corresponding potential profile is shown in the inset of Fig.~\ref{Fig1}(a). 
Our main results are qualitatively insensitive to the detailed barrier shape; this point is discussed in Appendix~\ref{shapes}.

\subsection{Analytical solutions of the GPE}
\label{Sec.analytical}

To investigate the stability properties of the BEC, we rely on analytical solutions of Eq.~(\ref{1DGPERes}). 
For a single static square barrier in an infinite 1D geometry, exact solutions are known in terms of Jacobi elliptic functions~\cite{Baratoff1970, Piazza2010}. 
In this case, the density profile inside the potential barrier takes the form
\begingroup
  \setlength{\jot}{6pt}
  \begin{equation}\label{solutionsins}
    \rho_{\mathrm{in}}(\theta) =
    \begin{cases}
      \rho_{0} + \rho_{1}\dfrac{s^{2}(\alpha_{1}\theta,m_{1})}{c^{2}(\alpha_{1}\theta,m_{1})},
        & \text{if }\Delta \ge 0 \text{ and } \rho_{1}\ge0,\\[\jot]
      \rho_{0} + \rho_{2}\dfrac{1-c(\alpha_{2}\theta,m_{2})}{1+c(\alpha_{2}\theta,m_{2})},
        & \text{otherwise},
    \end{cases}
  \end{equation}
\endgroup
where $s(z,m)$ and $c(z,m)$ are Jacobi sine and cosine functions, with argument $z = \alpha_i \theta$ and elliptic modulus $k_{i} = \sqrt{m_{i}}$, and $\rho_{0}$ is the density at the center of the barrier. 
The density profile outside the potential barrier, where the external potential of Eq.~(\ref{Barriers}) vanishes, is
\begin{equation}
    \label{solutionsout}
    \rho_{\mathrm{out}}(\theta)
    = \frac{J^2}{g\,\rho_{b}^2}
      + c_{b}\tanh^{2}{\biggl(\sqrt{1-\frac{2\xi_b^2J^{2}}{\rho_{b}^{2}}}\,
      \frac{\theta-\sigma}{\sqrt{2}\,\xi_b}+\phi_{0}\biggr)}^{2}.
\end{equation}
Here, $\rho_{b}$ is the bulk density, $\xi_b = 1/\sqrt{2g\rho_{b}}$ is the bulk healing length; the coefficients entering Eqs.~(\ref{solutionsins}) and~(\ref{solutionsout}) are reported in Ref.~\cite{footnotecoeff}.
As shown in Refs.~\cite{Baratoff1970, Piazza2010}, Eqs.~(\ref{solutionsins}) and~(\ref{solutionsout}) admit two distinct, non-degenerate pairs of solutions, corresponding to two different values of $\rho_0$. 
One solution corresponds to the lowest-energy configuration (ground state), while the other represents a higher-energy configuration (excited state). For the infinite system of Ref.~\cite{Piazza2010}, in the absence of a potential barrier the ground state reduces to a plane wave and the excited state to a dark soliton.

We use Eqs.~(\ref{solutionsins}) and~(\ref{solutionsout}) as approximate solutions for our 1D ring with $n$ rotating barriers, provided that the angular distance between neighboring barriers, $\theta_{j+1}-\theta_j$, is much larger than both the bulk healing length $\xi_b$ and the barrier width $\sigma$.
Under this condition, a bulk region with constant density $\rho_b$ can be defined. 
A key difference from the infinite 1D line considered in Refs.~\cite{Baratoff1970, Piazza2010} is that, in our toroidal geometry, the parameters $\rho_b$, $J$, and $\mu$ cannot be imposed externally: they must be determined self-consistently from the quantization condition Eq.~(\ref{Quant}) and, for finite $\omega$, by numerically solving Eq.~(\ref{1DGPERes}). 
A major advantage of the analytical approach is that it gives access to the excited state of the rotating system, which in turn provides information on the stability of the condensate at finite $\omega$.

Figures~\ref{Fig2}(a) and~\ref{Fig2}(b) show the  density profiles of the ground and excited states obtained for $n = 1$ and $n = 4$, respectively, as a function of $\omega$. 
In both cases, the density reaches its minimum at the center of the barrier, where the external potential (indicated by gray shaded areas) suppresses the condensate density. 
The numerical ground-state solution (circles) is in excellent agreement with the analytical expression given by Eqs.~(\ref{solutionsins}) and~(\ref{solutionsout}) (black solid line). 
Inside the barrier, the excited state (pink line) attains a lower density than the ground state, reflecting a larger kinetic energy. 
As $\omega$ approaches a critical value $\omega_{c}$, the two solutions merge. 
For $\omega>\omega_c$ no convergent numerical solution of the GPE is found. 
As confirmed below, by studying the evolution of the system following a small quench of $\omega$, the critical value $\omega_{c}$ marks the onset of a dynamical instability, signaled by the spontaneous emission of solitons in the 1D system.
We emphasize here that the notion of critical frequency $\omega_c$ is different from that of critical current $J_c$ studied for a superfluid in a necklace potential in Ref.~\cite{PezzeNATCOMM2024}.
The former is associated with a dynamical instability, while the latter is the largest value of the current reached in the system.
While in the weak barrier regime ($V_0/\mu \lesssim 1$), the maximum current is reached at $\omega=\omega_c$, in the deep Josephson regime ($V_0/\mu\gg1$), the current decreases before the system reaches the instability (here, $\sigma/\xi_b\approx 1$).
Further details are provided in Appendix~\ref{CurrApp}. \\
To further characterize the stability phenomenon, Fig.~\ref{Fig2}(c) displays the mean-field energy
\begin{equation}\label{GPEEN}
E = \int_{0}^{2\pi} d\theta\,\biggl(
    -\dfrac{1}{2}\sqrt{\rho}\,\dfrac{\partial^{2}\sqrt{\rho}}{\partial\theta^{2}}
    + \sqrt{\rho}\,V\sqrt{\rho}
    + \dfrac{g}{2}\rho^{2}
    + \dfrac{1}{2}\dfrac{J^{2}}{\rho}
\biggr)
\end{equation}
of the ground (solid lines) and excited (dashed lines) states as a function of the rotation frequency $\omega$ for different numbers of barriers $n$. 
At $\omega_{c}$, the energy of the ground state becomes degenerate with that of the excited state, in analogy with bifurcation processes~\cite{Hakim, Munoz-MateoPRA2015, SyafwanPhys2016}.
We emphasize that, according to Eq.~(\ref{Eq.omega}), the effective critical frequency can be accessed by changing the circulation $\nu$ and/or the rotation angular velocity $\Omega$.

The behavior of $\omega_{c}$ as a function of $n$ is shown in Fig.~\ref{Fig2}(d). 
The effective critical frequency increases monotonically with the number of barriers, $n$, as already suggested qualitatively by Fig.~\ref{Fig2}(c).
This monotonic dependence shows that the system becomes progressively more resilient to dynamical instabilities as $n$ increases, which is one of the main results of this work. 
For small $n$, when the density modulations associated with neighboring barriers do not significantly overlap, the dependence of $\omega_c$ on $n$ is approximately linear, as clarified in the following section.

\subsection{Increased stability with the number of barriers}\label{Increasing}

The mechanism underlying the enhanced stabilization can be understood in terms of two basic principles: 
particle-number conservation, enforced by density normalization, and circulation quantization, imposed by Eq.~(\ref{Quant}).
In general, the density profile $\rho(\theta)$ obtained from Eq.~(\ref{1DGPERes}) depends on both the effective rotation frequency $\omega$ and the number of barriers $n$. 
We make this dependence explicit by writing $\rho(\theta,\omega,n)$.
As shown in Fig.~\ref{Fig1}, under the condition of well-separated barriers, $\theta_{j+1}-\theta_j \gg \xi_b,~\sigma$, the density profile consists of a uniform bulk density and a set of identical depleted regions localized at the barriers.
Density normalization gives
\begin{equation}\label{Anal1}
\resizebox{0.97\linewidth}{!}{$
    \int_{0}^{2\pi} \rho(\theta,\omega,n)\, d\theta 
    = \rho_{b}(\omega,n)\,[2\pi - 2\tilde{\sigma} n]
    + n \int_{-\tilde{\sigma}}^{+\tilde{\sigma}}\rho(\theta,\omega,n)\, d\theta 
    = 1,$}
\end{equation}
where $\tilde{\sigma}$ is the characteristic length over which the density rises from its minimum at the barrier to the bulk value $\rho_{b}(\omega,n)$.
When $\xi_{b} \ll \sigma$, this length is on the order of the barrier half-width, $\tilde{\sigma} \approx \sigma$.
Equation~(\ref{Anal1}) leads to
\begin{equation}\label{Anal2}
\rho_{b}(\omega,n) =
\dfrac{1}{2\pi}\,
\dfrac{1 - n\displaystyle\int_{-\tilde{\sigma}}^{+\tilde{\sigma}}\rho(\theta,\omega,n)\, d\theta}
      {1 - n\,\tilde{\sigma}/\pi}.
\end{equation}
The integral in Eq.~(\ref{Anal2}) depends only weakly on $n$ and, since the density is depleted in the interval $[-\tilde{\sigma},\tilde{\sigma}]$, it is always smaller than $\tilde{\sigma}/\pi$. 
Equation~(\ref{Anal2}) therefore predicts an increase of $\rho_b$ with $n$.

Next, we analyze how increasing $n$ affects the superfluid phase, $\phi(\theta)$, along the ring. 
Using the quantization of circulation, we introduce the phase jump across a single barrier as
\begin{equation}\label{deltaphi}
    \delta\phi(\omega,n)
    = \dfrac{1}{n} \int_{0}^{2\pi} [\upsilon(\theta,\omega,n) - \upsilon_b]\, d\theta,
\end{equation}
where $\upsilon(\theta,\omega,n) = J/\rho(\theta,\omega,n) + \Omega$ is the superfluid velocity and $\upsilon_b = J/\rho_b + \Omega$ is its bulk value.
Using Eq.~(\ref{J2}), we obtain
\begin{equation}\label{deltaphi2}
    \delta\phi(\omega,n) = \dfrac{2\pi\,\omega}{I(\omega,n)^{-1} + n},
\end{equation}
where
\begin{equation}\label{integral}
I(\omega,n) = \int_{-\tilde{\sigma}}^{+\tilde{\sigma}}
\biggl(\frac{\rho_b(\omega,n)}{\rho(\theta,\omega,n)} - 1\biggr)\, d\theta.
\end{equation}
The integral in Eq.~(\ref{integral}) encodes the relative density depletion inside the barrier,
$[\rho_b(\omega,n)-\rho(\theta,\omega,n)]/\rho(\theta,\omega,n)$, and is expected to depend only weakly on $n$.
Neglecting this dependence, Eq.~(\ref{deltaphi2}) shows that $\delta\phi(\omega,n)$ decreases with increasing $n$ at fixed $\omega$~\cite{PezzeNATCOMM2024}. 
A detailed numerical analysis (see Appendix~ \ref{AppPhase} ) confirms this trend.

The instability sets in when the phase jump across each barrier reaches a critical value $\delta\phi_c$, beyond which phase slips are triggered. 
Since $\delta\phi(\omega,n)$ decreases with $n$ at fixed $\omega$, increasing the number of barriers moves the system away from the phase-slip threshold. 
This provides a qualitative explanation of the enhanced stability observed in Fig.~\ref{Fig1}(a) and Fig.~\ref{Fig2}(d).

Focusing now on the critical frequency, Eq.~(\ref{deltaphi2}) gives
\begin{equation}\label{critfreq2}
    \omega_c(n) = \dfrac{\delta\phi_c(n)}{2\pi}
    \biggl(\dfrac{1}{I_c(n)} + n\biggr),
\end{equation}
where $\delta\phi_c(n) \equiv \delta\phi(\omega=\omega_c,n)$ and $I_c(n) \equiv I(\omega=\omega_c,n)$ denote the values at the critical point.
Equation~(\ref{critfreq2}) is a nonlinear implicit relation for $\omega_c(n)$. 
However, we find numerically that both $\delta\phi_c(n)$ and $I_c(n)$ vary only weakly with $n$ (see Appendix \ref{AppPhase}).
This implies that $\omega_c(n) \propto n$ .

A further simplification holds if $\delta\phi_c \approx \pi/2$, which is approximately true for $V_0/\mu \approx 1$ and narrow barriers, $\sigma/\xi \approx 1$.
Neglecting again the dependence of $I_c(n)$ on $n$, we obtain
\begin{equation}\label{critfreq3}
    \omega_c(n) \approx \omega_0 + \dfrac{n}{4},
\end{equation}
with $\omega_0 = 1/(4 I_c)$.
Equation~(\ref{critfreq3}) predicts that the slope of $\omega_c(n)$ as a function of $n$ is equal to $1/4$, independent of the other system's parameters.
In Appendix \ref{AppPhase}, we compare this prediction with the numerical results and find an overall good agreement, with deviations due to the approximations discussed above.

\begin{figure}[t]
\centering
\includegraphics[width=\columnwidth]{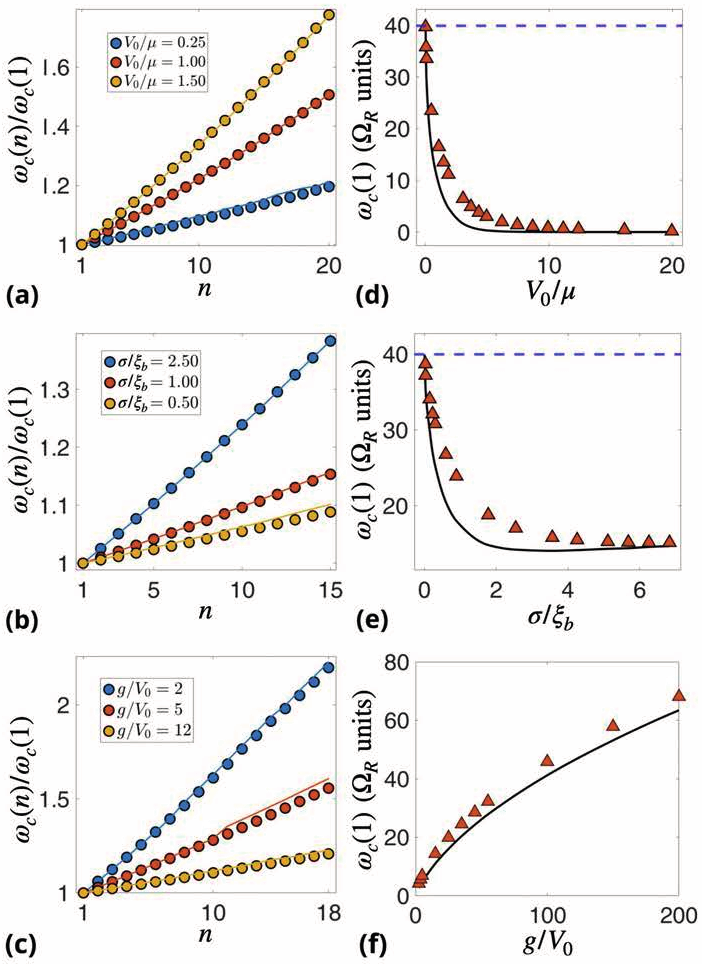}
\caption{
Critical effective rotation frequency $\omega_{c}(n)$ as a function of the number of barriers $n$ for other system parameters. 
In panels~\ref{Fig3}(a)–(c), the critical frequency is normalized to its value at $n=1$, i.e., $\omega_c(n)/\omega_c(1)$. 
We vary the barrier height $V_{0}/\mu$ in (a), the barrier width $\sigma/\xi_b$ in (b), and the coupling constant $g$ (reporting the ratio $g/V_0$) in (c). 
Dots show the numerical solutions of Eq.~(\ref{1DGPERes}), while solid lines correspond to the prediction of Eq.~(\ref{critfreq2}).  
Panels~\ref{Fig3}(d)–(f) display the critical frequency at a single barrier, $\omega_c(1)$, as a function of $V_{0}/\mu$ (d), $\sigma/\xi_b$ (e), and $g/V_0$ (f) (orange triangles). 
The black solid line is Eq.~(\ref{instcrit}). Parameters are: (a) $g=5000$ and $\sigma/\xi_b\simeq 1.2$; (b) $g=10^4$ and $V_0/\mu\simeq 0.7$; (c) $\sigma/\xi_b\simeq 0.6$ and $V_0/\mu\simeq 2.2$; (d) $g=10^4$ and $\sigma/\xi_b\simeq 0.9$; (e) $g=10^4$ and $V_0/\mu\simeq 0.12$; (f) $V_0/\mu\simeq 0.12$ and $\sigma/\xi_b\simeq 0.9$.}
\label{Fig3}
\end{figure}

\begin{figure*}[t!]
\centering
\includegraphics[width=1\textwidth]{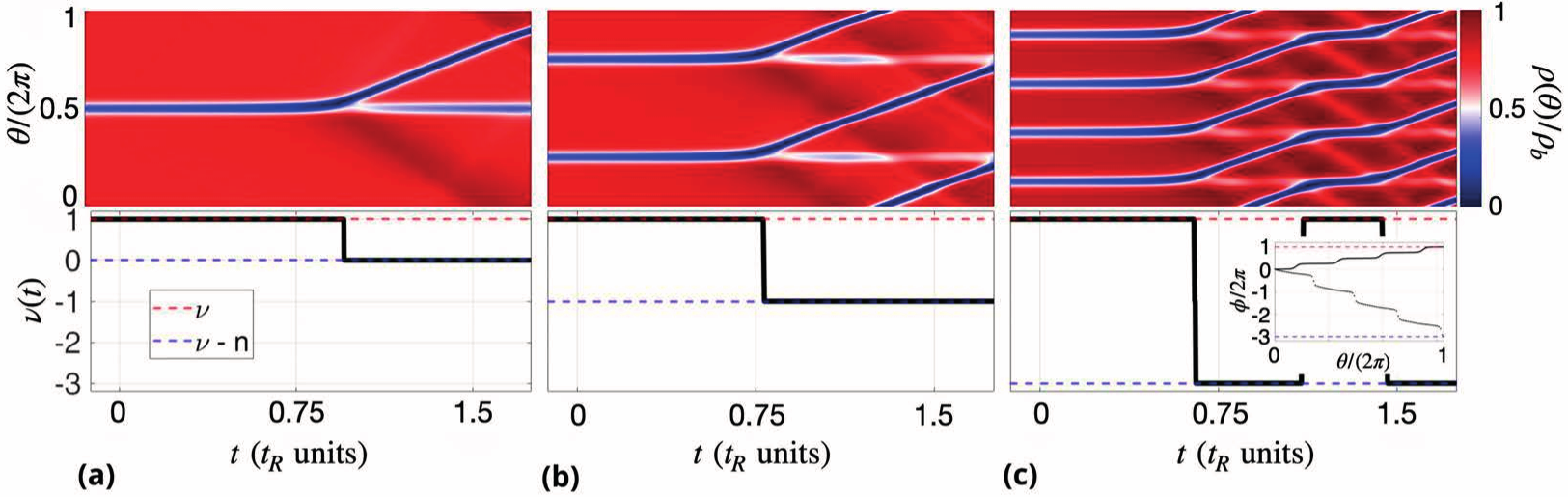}
\caption{
Joint emission of solitons in the clean necklace.
Each panel shows the time evolution of the density profile (top) and the winding number $\nu(t)$ (bottom).
Different panels correspond to different numbers of barriers: $n=1$ (a), $n=2$ (b), and $n=4$ (c).
In each case, the rotation frequency is quenched to a value $\omega = \omega_c(n) + \delta\omega$, with $\delta\omega \simeq 0.05$.
In the lower panels, the red dashed line indicates the initial winding number $\nu(0) = 1$, while the blue dashed line marks the value $\nu = \nu(0) - n$ observed immediately after soliton emission.
The inset in panel (c) shows the superfluid phase as a function of $\theta$ at different times: $t=0.5$ (solid line), corresponding to a circulation $\nu=1$, and $t=1$ (dotted line), corresponding to a circulation $\nu=-3$.
Parameters are the same as in Fig.~\ref{Fig2}.
}
\label{Fig4}
\end{figure*}

\begin{figure*}[t]
\centering
\includegraphics[width=0.9\textwidth]{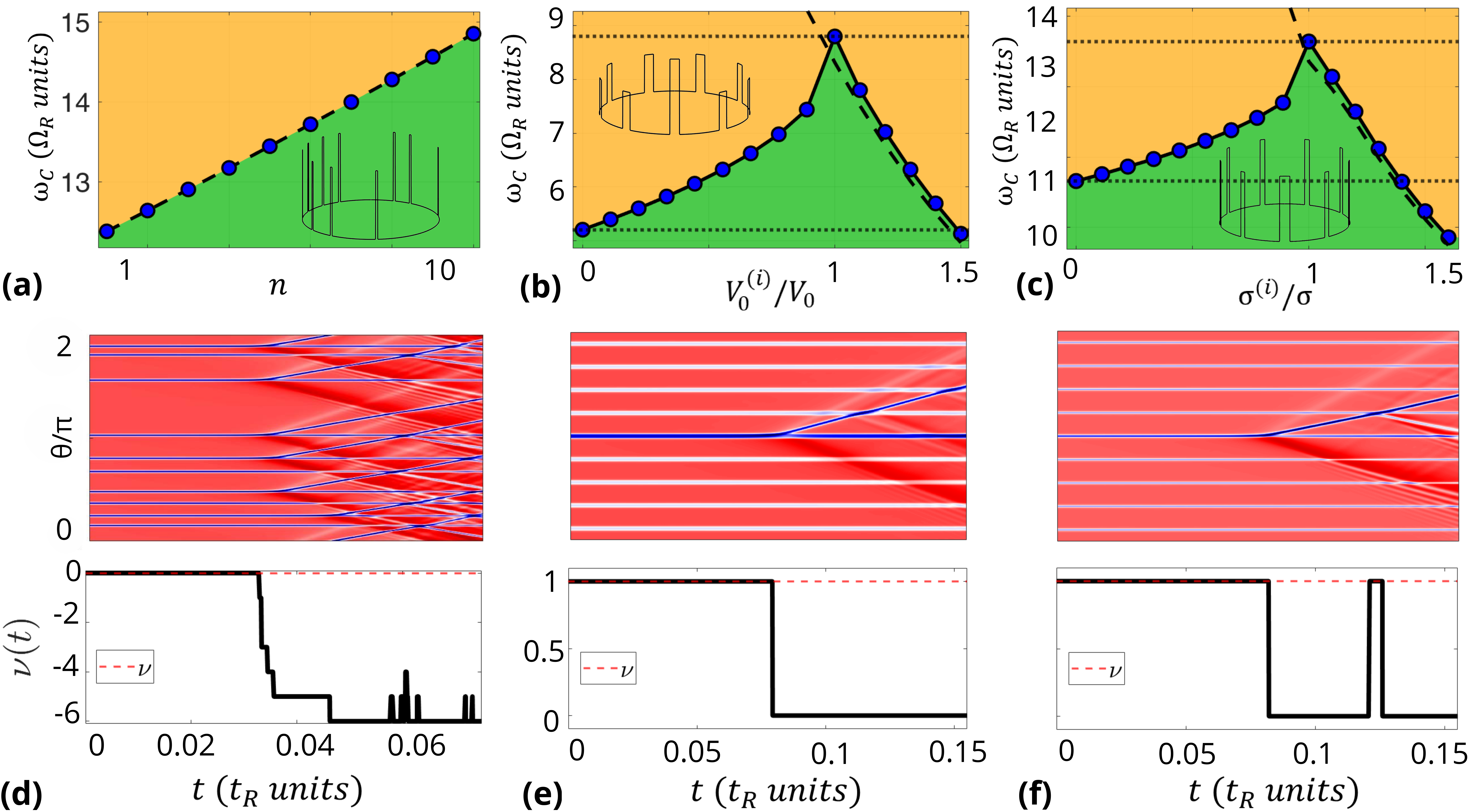}
\caption{
Panels (a)–(c) show the critical frequency in the dirty necklace with non-uniform square barriers; examples of the corresponding potentials are displayed in the insets.
Panels (d)–(f) present the associated density dynamics (top) and winding number (bottom) as a function of time.
In all cases, the dynamics is triggered by quenching the rotation frequency to $\omega = \omega_c(n) + \delta\omega$, with $\delta\omega \simeq 0.05$.
(a), (d): randomly distributed barriers (see inset).
Panel (a) shows $\omega_c(n)$ (dots) as a function of $n$, obtained by averaging over 30 random realizations of the potential (the rms fluctuations are smaller than the symbol size).
The dashed line corresponds to a uniform necklace with the same $n$, $V_0$, and $\sigma$.
Here, $g = 10^4$, $V_0/\mu \sim 1$, $\sigma/\xi \sim 1$, and $\nu(0) = 0$.
(b), (e): non-uniform barrier height.
One barrier has height $V_0$, while the remaining $n-1$ barriers have height $V_0^{(i)}$.
Panel (b) shows the critical frequency as a function of $V_0^{(i)}$.
(c), (f): non-uniform barrier width.
One barrier has width $\sigma$, while the remaining $n-1$ barriers have width $\sigma^{(i)}$.
Panel (c) reports the critical frequency as a function of $\sigma^{(i)}$.
In panels (b) and (c), the upper dotted line indicates $\omega_c$ for $n$ identical barriers, while the lower dotted line gives $\omega_c$ for a single barrier ($n=1$). Dashed line indicates $\omega_c$ for the configuration of $n = 8$ barriers.
In panels (b) and (e), the parameters are $n=9$, $g = 5000$, $V_0/\mu \sim 0.6$, $\sigma/\xi_b \sim 6$, and $\nu(0) = 1$.
In panels (c) and (f), the parameters are $n=9$, $g = 5000$, $V_0/\mu \sim 0.6$, $\sigma/\xi_b \sim 1.5$, and $\nu(0) = 1$.
}
\label{Fig5}
\end{figure*}

\begin{figure*}[t]
\centering
\includegraphics[width=1\textwidth]{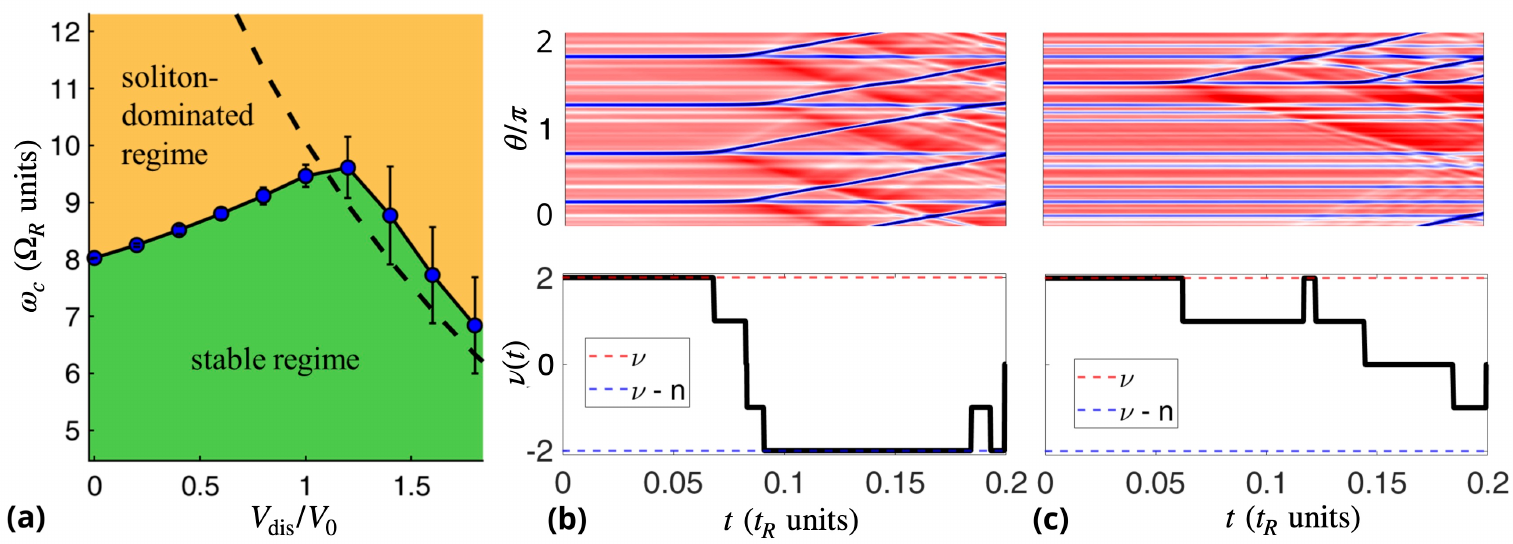}
\caption{
(a) Critical frequency $\omega_c$ in the dirty necklace (blue dots), as a function of $V_{dis}/V_0$ and for $n=4$.
Data are obtained by averaging over $35$ different realizations of the speckle potential, with error bars giving the standard deviation.
The black line is a guide to the eye, while the black dashed line is $\omega_c$ for the sole speckle disorder. 
As in Fig.~\ref{Fig1}(b), the green (yellow) shading highlights the stable (unstable) region.
Panels (b) and (c) show the density  (top) and the corresponding winding number $\nu(t)$ (bottom) as a function of time for $V_{dis} = 0.6V_0$ and $V_{dis} = 1.4V_0$, respectively. 
The dynamics is triggered by a quench to $\omega = \omega_c + \delta\omega$, with $\delta\omega \approx 0.01$. 
Parameters are $g = 3000$, $V_0/\mu \sim 0.5$, $\sigma/\xi_b \sim2$ and $\sigma/\gamma \sim 2$, where $\gamma$ is the disordered potential's correlation length.
}
\label{Fig6}
\end{figure*}

In Fig.~\ref{Fig3}(a)–(c), we show the results of numerical simulations for $\omega_c(n)$ as a function of $n$ (dots), obtained by varying the barrier height $V_{0}$, the width $\sigma$, and the coupling constant $g$, respectively.
Solid lines are the prediction of Eq.~(\ref{critfreq2})~\cite{note}. 
The stabilization effect is present both in the weak-link ($V_{0}/\mu < 1$) and in the Josephson ($V_{0}/\mu > 1$) regimes. 
In particular, the slope of $\omega_c(n)$ increases with $V_{0}/\mu$. 
We also explore a wide range of barrier widths, covering both the hydrodynamic regime ($\sigma/\xi_b > 1$) and the thin-barrier regime ($\sigma/\xi_b < 1$), and find that the slope of $\omega_c(n)$ increases with $\sigma/\xi_b$.

Overall, enhancing the barrier effect (e.g., by increasing $V_{0}/\mu$ or $\sigma/\xi_b$, or by decreasing $g$) increases the separation between the critical frequencies at successive $n$, thereby making the stabilization mechanism more pronounced, i.e., yielding a larger slope of $\omega_c(n)$. 
Conversely, for very strong interactions (large $g$) or vanishingly weak and narrow barriers (small $V_{0}/\mu$ and $\sigma/\xi_b$), the stabilization effect is still observed, but with very weak increase of $\omega_c(n)$ with $n$. 

Our analysis confirms that the enhancement of rotational stability is primarily controlled by the number of barriers, rather than by microscopic details of the potential. 

In Fig.~\ref{Fig3}(a)–(c), $\omega_c(n)$ is normalized to the corresponding value at $n=1$, $\omega_c(1)$. 
For completeness, Figs.~\ref{Fig3}(d)–(f) show $\omega_c(1)$ as a function of $V_{0}/\mu$, $\sigma/\xi_b$, and $g$, respectively. 
Increasing $V_{0}$ (at fixed $g$ and $\sigma/\xi_b \approx 1.3$) leads to a decrease of $\omega_c(1)$; a similar trend is observed when increasing $\sigma$ at constant $V_{0}$ and $g$. 
In particular, in the limit $g \to 0$ we find that the slope of $\omega_c(n)$ as a function of $n$ increases [panel (c)], while $\omega_c(1)$ tend to zero [panel (f)].
\subsection{Comparison with the Landau instability criterion}

As discussed above, the critical frequency $\omega_c$ marks the onset of a dynamical instability in the rotating system.
It is therefore natural to compare our results with the Landau instability criterion, derived in Ref.~\cite{Watanabe} for a 1D superfluid with a single potential barrier.
In that framework, the instability is expected when the superfluid velocity in the frame co-moving with the barrier, $J/\rho(\theta)$, equals the local sound speed $c_s(\theta)=\sqrt{g\rho(\theta)}$.
This condition is first met at the barrier center (at $\theta=\theta_j$), where the density is minimal and the velocity is maximal, according to Eq.~(\ref{J}).
Imposing $J/\rho_0 = \sqrt{g\rho_0}$, where $\rho_0$ is the density at the center of the barrier, gives
\begin{equation}\label{instcrit}
\omega_c = \dfrac{2\pi\sqrt{g\rho_{0}^3}}{f_s}.
\end{equation}
In Figs.~\ref{Fig3}(d)–(f) we compare the prediction of Eq.~(\ref{instcrit}) (black solid lines) with the numerical data for a single barrier (orange triangles).
Our results are consistent with previous theoretical studies~\cite{Hakim,Watanabe}, which show that the critical velocity associated with a Landau instability decreases as the barrier strength is increased (either by raising the barrier height $V_0$ or its half-width $\sigma$).
Conversely, when the barrier becomes weaker, the critical frequency approaches the bulk sound velocity $\sqrt{g\rho_b}$, as expected for a nearly uniform superfluid~\cite{Hakim,Watanabe}.
Since the sound speed increases with $g$, stronger interactions enhance the robustness of the system against perturbations, leading to higher critical frequencies.

Equation~(\ref{instcrit}), originally derived in the hydrodynamic regime~\cite{Watanabe}, also agrees well with our simulations in the thin-barrier limit ($\sigma/\xi_b \ll 1$) and for weak barriers ($V_0/\mu < 1$).
In the intermediate regime $\sigma \sim \xi_b$, however, Eq.~(\ref{instcrit}) exhibits systematic deviations from the numerical data, although it reproduces the qualitative dependence on $\sigma/\xi_b$.
In this case, the condition $J/\rho = c_s$ is not fulfilled exactly at the barrier center but rather within a distance of order $\xi_b$ from it, so that the Landau critical velocity corresponds to a fraction of the bulk sound speed $\sqrt{g\rho_b}$, larger than the minimum value $\sqrt{g\rho_0}$.

\subsection{Soliton emission and inversion of the current}

We now explore the behavior of the system for $\omega>\omega_c$ -- the regime indicated as ``soliton-dominated'' in Fig.~\ref{Fig1}(a).
In this regime, the winding number $\nu$ depends on time, as it may change during the dynamics.
We perform real-time simulations of Eq.~(\ref{1DGPEFULL}). 
The superfluid is initially prepared in the stable regime, with $\omega < \omega_c$.
At time $t=0$, the rotation frequency is quenched to a final value $\omega + \Delta\omega > \omega_c$.
Here we consider $\Delta \omega \ll \omega_c$ in order to limit spurious excitations.

After the quench, the system becomes dynamically unstable and, as suggested by the discussion in Sec.~\ref{Sec.analytical}, it relaxes by emitting solitons.
The ensuing dynamics is particularly rich:  representative examples are shown in Fig.~\ref{Fig4}(a)–(c).
In the upper panels, we plot the co-rotating density profile $\rho(\theta)$ for $n=1,2,4$, respectively, as a function of time.
Initially, the density exhibits a dip at the position of each barrier; in the co-rotating frame the barriers are at rest.
After a time of order a fraction of $t_R = mR^2/\hbar$, we observe the simultaneous emission of $n$ dark solitons, one from each barrier.
Once formed, the solitons propagate along the ring, as visible from the slanted density dips.
The density remains depleted at the barrier positions, and sound waves are emitted, propagating in the direction opposite to the soliton motion.
As clearly visible in panel (c), solitons may remain pinned to a barrier for some time before being re-emitted.

The lower panels of Fig.~\ref{Fig4}(a)–(c) show the circulation $\nu(t) = \tfrac{\phi(2\pi,t)-\phi(0,t)}{2\pi}$ [see Eq.~(\ref{Quant})] as a function of time.
We observe sudden jumps in $\nu(t)$ when solitons are emitted from the barriers; the jump is quantized and equal to the number of emitted solitons.
This phase-slip process has been observed in similar regimes involving quantized vortices~\cite{Xhani_2020,PiazzaPhys2009,WrightPRL2013}.
After the joint soliton emission, the winding number takes the value $\nu(t) = \nu(0) - n$ (dashed blue line), where $\nu(0)$ is the initial circulation (dashed red line, $\nu=1$ in the figure) and $n$ is the number of barriers.
A particularly interesting scenario occurs for $\nu(0)=1$ and $n=2$ [Fig.~\ref{Fig4}(b)], where the winding number reaches $\nu(t)=-1$ for sufficiently long evolution times, effectively realizing a switch that reverses the circulation.
For $\nu(0)=1$ and $n=4$ [Fig.~\ref{Fig4}(c)], soliton emission not only reverses the sign of the circulation but also drives it to $\nu(t)=-3$, i.e., below $-\nu(0)$.
When solitons pass through the barrier region, they can be temporarily absorbed, during which interval the winding number returns to its initial value $\nu(0)$.
In the inset of Fig.~\ref{Fig4}(c) we show the superfluid phase at time $t=0.5$ (solid line), when the circulation is equal to $\nu(0)$, and at $t=1$ (dotted line), when the circulation has dropped by $n$ units. 
Overall, these results demonstrate that, by tuning the number of barriers in the clean necklace, one can control the phase-slip dynamics and, consequently, the circulation in the ring.

Finally, we recall that the above effects hold for $g>0$. In the case of attractive interaction $g<0$, we still observe an increase of the critical frequency $\omega_c$ with $n$.
Yet, the instability for overcritical rotations is qualitatively different: we observe oscillations of density peaks and associated phase slips, giving staggered dynamical changes of the winding number $\nu(t)$.
A detailed analysis is reported in Appendix~\ref{Att}.

\section{Dirty necklace}\label{DirtySec}

In the previous analysis all barriers were identical and equally spaced, characterized by the same height $V_0$ and width $\sigma$.
We now relax this condition and refer to the case of a non-uniform ring potential as {\it dirty necklace}.

\subsection{
Non-uniform square barrier potentials
}

We first consider the case
\begin{equation}\label{Barriers2}
    V(\theta) = 
    \begin{cases}
      V_{0}^{(i)}, & \text{for } \theta \in [\theta_{i} - \sigma^{(i)}, \theta_{i} + \sigma^{(i)}],\\
      0,           & \text{otherwise},
    \end{cases}
\end{equation}
where the superscript emphasizes that these parameters may differ from barrier to barrier.
In Fig.~\ref{Fig5}(a) we plot $\omega_c(n)$ as a function of $n$ (dots) for a configuration in which the barrier centers $\theta_i$ are randomly sampled under the constraint $\theta_{i+1}-\theta_i \geq 2\xi_b$ [see inset of Fig.~\ref{Fig5}(a)].
Otherwise, all barriers have the same height and width, $V_i = V_0$ and $\sigma_i = \sigma$.
This condition guarantees the existence of regions with well-defined bulk density $\rho_b$, consistent with the discussion above.
Although spatial periodicity is lost, the stabilization effect persists: the critical frequency $\omega_c$ still increases almost linearly with $n$, and the numerical data for random configurations essentially coincide with the uniform case (black dashed line).
This robustness arises because, in this setting, the clean and dirty necklaces share the same phase drop at each barrier.
In contrast, the soliton dynamics following a quench of $\omega$ differs from the periodic case, as shown in Fig.~\ref{Fig5}(d) for $n = 10$.
Solitons are nucleated at different times, and within the simulated time window only a subset of the barriers emits a soliton.
As a consequence, $\nu(t)\neq \nu(0) - n$, since the number of emitted solitons is smaller than $n$.
Nevertheless, even in this situation it is possible to invert the circulation.

A different scenario arises when the barriers remain equally spaced but one of them differs in height or width from the others.
These cases are shown in Figs.~\ref{Fig5}(b) and~(c), respectively.
Specifically, one barrier of height $V_0$ and half-width $\sigma$, and the other $n-1$ barriers with height $V_0^{(i)}$ and half-width $\sigma^{(i)}$.
The case $V_0^{(i)}=0$ or $\sigma^{(i)}=0$ corresponds to the single-barrier case, while $V_0^{(i)}=V_0$ and $\sigma^{(i)}=\sigma$ to the periodic arrangement.
The top dashed line in Figs.~\ref{Fig5}(b) and~(c) shows the uniform case with $n = 9$ identical periodic barriers, while the bottom dashed line indicates the critical frequency for $n = 1$.
Increasing $V_0^{(i)}$ or $\sigma^{(i)}$ increases $\omega_c(n)$ with respect to the single barrier case, reaching a maximum at the uniform values $V_0$ and $\sigma$. For larger values, $\omega_c$ progressively decreases.
Importantly, the reduction of $\omega_c(n)$ is not abrupt and the system with $n$ barriers can remain more stable than the single-barrier case in a relatively broad range.
The behavior is mainly governed by the overall properties of the barrier array, which determine how the phase drop $\delta\phi$ is shared across the ring.
The corresponding soliton dynamics, shown in Figs.~\ref{Fig5}(e) and~(f) for finite $V_0^{(i)}$ and $\sigma^{(i)}$, exhibits similar features to the non-periodic case.
When the rotation frequency is quenched above $\omega_c$, we typically observe the emission of a single soliton.
The associated phase slip changes $\nu(t)$ by $\pm 1$, adding or removing one quantum of circulation.
The soliton nucleates at the strongest barrier (largest $V_0^{(i)}$ or $\sigma^{(i)}$), where the density depletion is maximal.

Overall, these results indicate that the stabilization induced by $n-1$ barriers, compared to the single-barrier case, does not rely on perfect periodicity.
This motivates the study of a fully disordered configuration, where the $n$-barrier necklace is superimposed on a non-uniform background potential, as discussed in the next section.

Finally, we have also considered the case of each barrier rotating at a different frequency. 
Consistent with our earlier analysis, soliton emission is governed by the fastest barrier and occurs only when the local rotation exceeds the critical threshold, see Appendix~\ref{app3}.
This further reinforces the conclusion that stability is controlled by the strongest local perturbation.

\subsection{Combination of square barriers and speckle disorder}

We finally consider the case of background potential is placed between the barriers of Eq.~(\ref{Barriers}):
\begin{equation} \label{eq:V_total_speckle}
    {V}(\theta) = \begin{cases}
      V_{0}, &  \text{for } \theta \in [\theta_j - \sigma, \theta_j + \sigma],\\
      V_{\mathrm{sp}}(\theta), &  \text{elsewhere}, 
    \end{cases}
\end{equation}
where $\theta_j = 2\pi j/n$ and $V_{\rm sp}(\theta)$ is a disordered speckle potential~\cite{Sanchez-PalenciaPRL2007, BillyNATURE2007, PezzePRL2011, notedisorder} of maximum height $V_{dis}$ and correlation length $\gamma$.

In Fig.~\ref{Fig6}(a) we show the critical frequency $\omega_c(n)$ as a function of the disorder amplitude $V_{\rm dis}$ for $n = 4$.
Each point is obtained by averaging over $35$ independent realizations of the speckle potential.
We find that $\omega_c$ increases with $V_{\rm dis}$ and reaches a maximum around $V_{\rm dis} \sim V_0$.
For stronger disorder, $V_{\rm dis} \gtrsim V_0$, the speckle potential dominates over the periodic barriers: the critical frequency decreases mainly due to the peaks of the disordered potential having random height.
The maximum of $\omega_c$ depends on the disorder correlation length $\gamma$: for smaller $\sigma/\gamma$ ratios (more correlated disorder) it shifts to smaller $V_{\rm dis}$, whereas for very short correlation lengths ($\sigma/\gamma\gg1$) $\omega_c$ may continue to increase even for $V_{\rm dis} > V_0$.

Overall, regardless of the specific parameters, we consistently observe an initial increase of $\omega_c$ with $V_{\rm dis}$, as sketched in Fig.~\ref{Fig1}(b): surprisingly, disorder enhances the stability of the system against dynamical excitations. 
This behavior is similar to the one observed in Fig.~\ref{Fig5}(b). 
We thus conclude that the stabilization effect factor observed for low disorder (and weak barrier height, for the ordered case) is attributed to the increase in the number of effective barriers rather than a pure randomness effect.

Figures~\ref{Fig6}(b) and (c) display the density profiles (top panels) and the winding number (bottom) as a function of time following a quench of $\omega$.
The dynamics is intermediate between the cases discussed above.
When $V_{\rm dis} \ll V_0$ [Fig.~\ref{Fig6}(b)], $n$ solitons are emitted, as in Fig.~\ref{Fig4}, although their emission is not strictly simultaneous.
The winding number decreases in steps of $n$, so that $\nu(t) = \nu(0) - n$ after emission.
In contrast, when $V_{\rm dis} \gtrsim V_0$ [Fig.~\ref{Fig6}(c)], solitons are emitted only from the strongest effective barriers, corresponding to the deepest density minima, as in Figs.~\ref{Fig5}(e) and~(f).
In this regime, $\nu(t)$ changes by one unit at a time, for each emitted soliton.

\section{Conclusions}

Our work, focused on a 1D BEC at zero temperature, has allowed us to systematically explore a wide range of parameters and to obtain analytical results regarding the enhanced robustness of the  superfluid necklace to dynamical excitations. 
Despite its simplicity, the model captures effects that are expected to persist in more complex systems and geometries.
In particular, we have shown that the stabilization mechanism induced by multiple barriers is present in both the hydrodynamic and tunneling regimes, with the slope of the critical effective rotation frequency $\omega_c(n)$ increasing with the barrier height and width.
Moreover, we have unveiled a rich dynamics in the  unstable regime: in the clean ring, solitons are emitted simultaneously from each barrier, enabling counterintuitive configurations in which the circulation can be controlled and even reversed by tuning $\omega$ and the number of barriers.
The stabilization effect has a robust topological nature that extends to dirty necklaces with imperfections and inhomogeneities.
Remarkably, a disordered potential of moderate amplitude can further enhance $\omega_c(n)$, offering a novel perspective on the subtle and counterintuitive interplay between superflow and impurities~\cite{NeverovCAMMPHYS2022, GastiasoroPRB2018, LerouxPNAS2019, KwokRMP2016}.
Beyond motivating new experimental investigations, our results call for advanced theoretical studies and extensions to Fermi superfluids~\cite{TuzemenARXIV} and supersolids~\cite{TengstrandPRA2021,SindikPRL2024,DonelliPRA2025,PretiARXIV} in toroidal traps, as well as for the inclusion of thermal and quantum fluctuations~\cite{AmicoPRL2005, SnizhkoPRA2016, KumarPRA2017, MehdiSCIPOST2021}.
In summary, our work demonstrates the possibility of exploiting multiply connected geometries to engineer both stable and unstable dynamics in ring superfluids over a broad range of parameters, with direct relevance for current experiments and atomtronic devices~\cite{SeamanPRA2007, AmicoRMP2022, PoloQST2024}.
 
\section{Acknowledgments}

We thank Marzena Ciszak, Giulia Del Pace, Beatrice Donelli, Nicola Grani, Diego Hernandez-Rajkov, Francesco Marino, Giacomo Roati, Francesco Scazza and Klejdja Xhani for discussions.
This work has been supported by the Horizon Europe programme HORIZON-CL4-2022-QUANTUM-02-SGA (project PASQuanS2.1, GA no. 101113690).

\section{Appendix}

\subsection{Attractive interactions} \label{Att}
\begin{figure}[t!]
\centering
\includegraphics[width=\columnwidth]{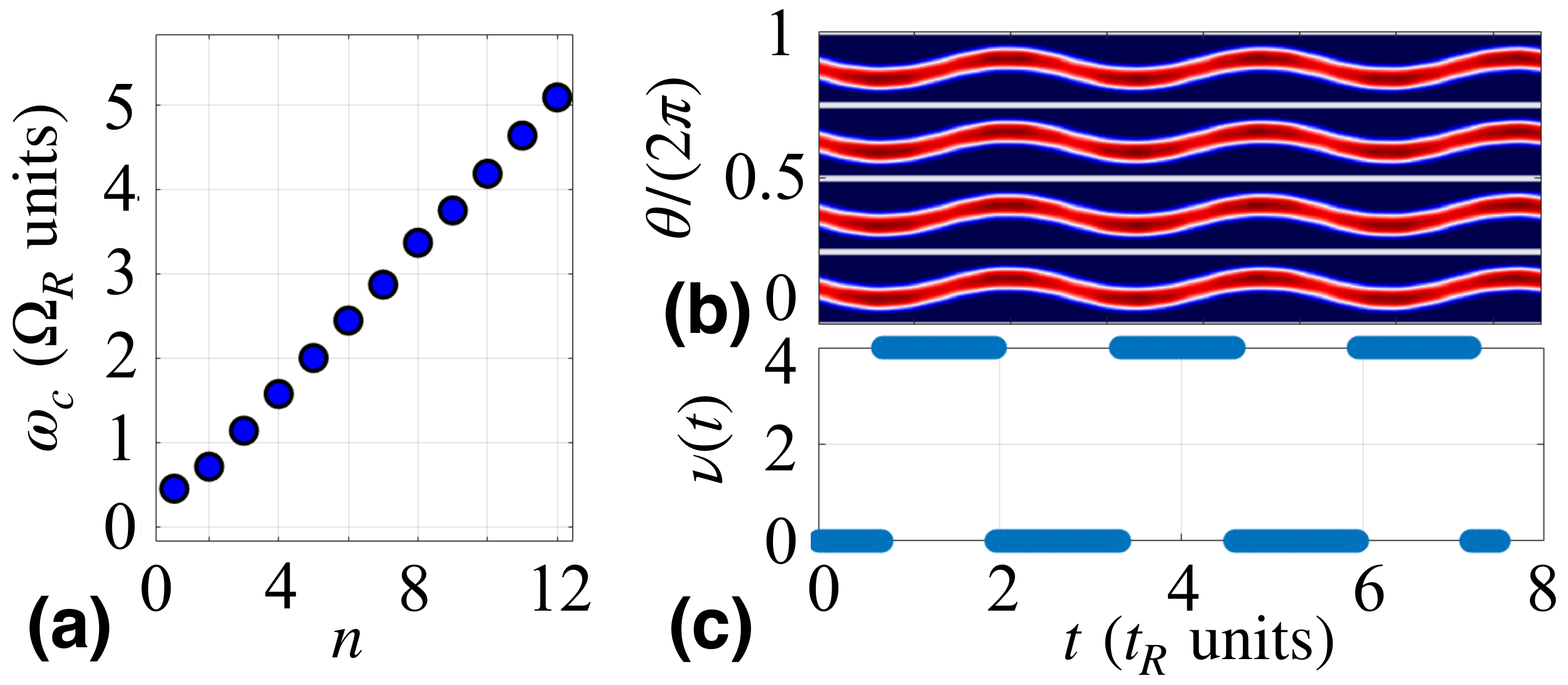}
\caption{
(a) Critical frequency $\omega_c$ as a function of $n$. (b) Time-evolution of the system's density after a dynamical quench. Grey lines report the position of the potential barriers. \textbf{(c)} Winding number $\nu$ as a function of time $t$. Parameters are set as $V_0/\mu \approx 10$, $g = -15$, $\sigma/\xi \sim 0.1$.
}
\label{SOLI}
\end{figure}

We now consider the case of weakly attractive interactions $g<0$ and $|g|<|g_{c}|$, where $g_{c}$ is a critical bifurcation value.
For $|g|>|g_{c}|$ the system develops a bright soliton localized at one of the potential minima~\cite{Salasnich2022}.
Differently from the case of repulsive interaction, where the density is almost uniform and drops across the barrier, in the case of attractive interactions the density is characterized by $n$ sharp peaks, located between each barrier.
Interestingly, upon scanning the rotation frequency $\omega$, we still find a critical threshold value for $\omega_c$.
As shown in Fig.~\ref{SOLI}(a), $\omega_c$ increases approximately linearly with the number of barriers $n$, as in the repulsive case.
If the rotation frequency is perturbed below $\omega_c$ (namely we prepare a stable ground state of frequency $\omega<\omega_c$ and quench it to values $\omega+\delta \omega<\omega_c$), then the system remains stable and we observe small oscillations of the density peaks. 
Conversely, when we quench the rotation frequency $\omega<\omega_c$ to values $\omega+\delta \omega>\omega_c$, we observe that the system undergoes an instability.
This manifests by phase slips corresponding to staggered dynamical changes in the winding number $\nu(t)$: we find that $\nu(t)$ jumps back and forth from $\nu(0)$ to values $\nu(0)+n$ following the oscillation of the density peak between the barriers. 
As shown in Fig.~\ref{SOLI}(b), the density peaks initially move (in the rotating frame) in the same direction as the change of angular frequency.
This happens until they approach the potential barriers: crossing the barrier is not possible because it is inhibited by the high energy required for a density drop. 
Therefore, density peaks are "repelled" by the barriers and move against the rotation of the system: this is only possible by a phase slip process.
As shown in Fig.~\ref{SOLI}(c), we observe the emission of exactly $n$ phase slips.
Again the density peaks, approaching the barrier, are repelled back, resulting in oscillations accompanied by phase slips. 
Finally, we have noticed that the critical interaction threshold $\vert g_c\vert$ also increases with the number of barriers $n$.
In fact, increasing $n$ determines a local decrease of density across the barrier and thus an increase of the density peaks, which makes the system more stable against the collapse to a single bright soliton.
This is compatible with the results of Ref.~\cite{Salasnich2022}, which predict an increase of $g_c$ in the presence of a single potential barrier in a 1D ring potential, when increasing the barrier's strength. 
In our case, we increase the number of barriers $n$, while the ratios $V_0/\mu$ and $\xi/\sigma$ are kept  constant.

\subsection{Current and critical frequency}\label{CurrApp}
\begin{figure}[t!]
\centering
\includegraphics[width=\columnwidth]{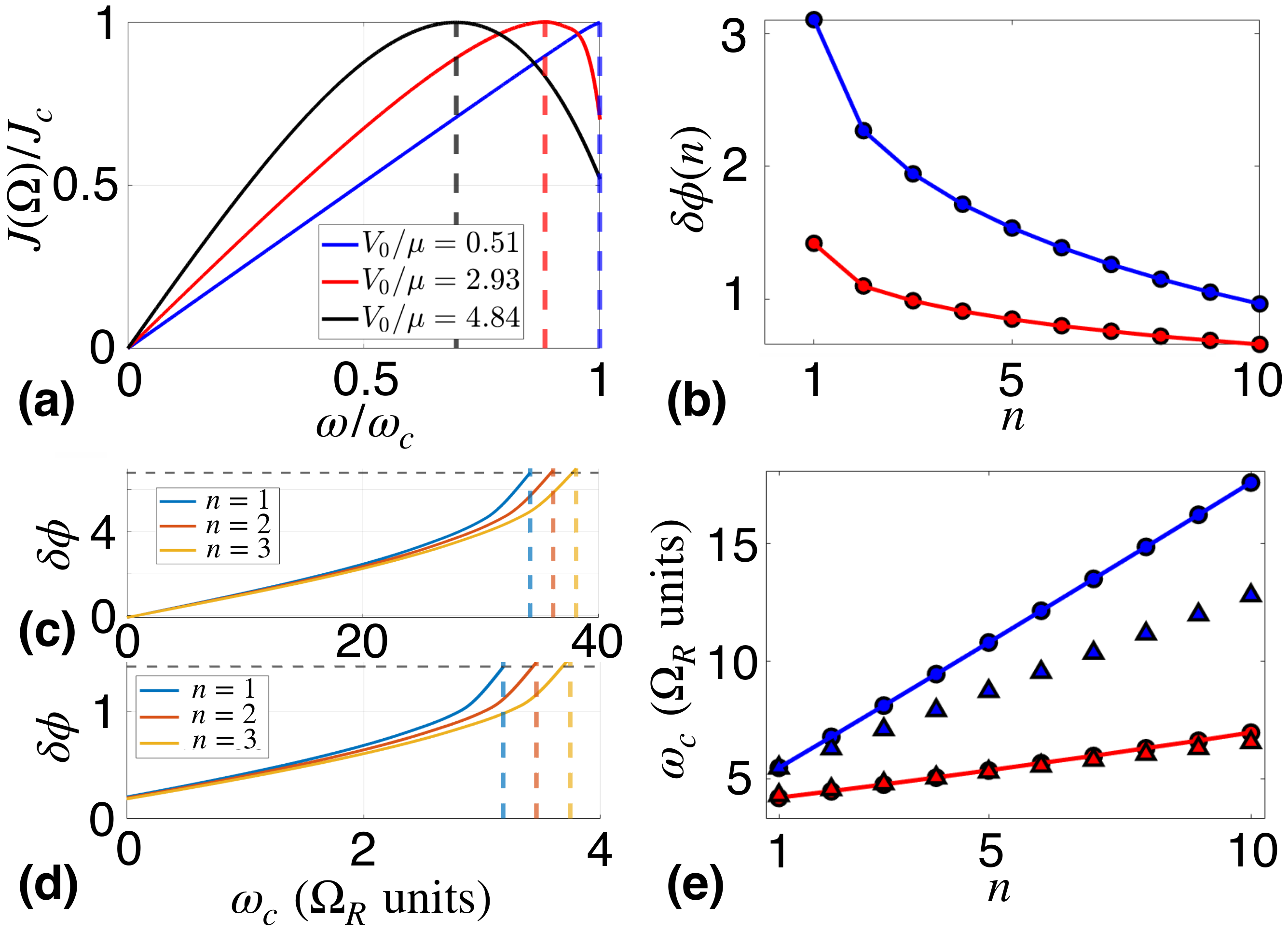}
\caption{(a) Superfluid current $J$ (rescaled to its maximum value $J_c$) for different values of $V_0/\mu$ as a function of the frequency ratio $\omega/\omega_c$. Parameters are set as $g = 3000$, $\sigma/\xi \sim 1$.
(b) Phase jump $\delta\phi$ as a function of the number of obstacles $n$. Red (Blue) data reports the Josephson (hydrodynamic) results. Full dots represent numerical results, while solid lines report the prediction of Eq.~(\ref{deltaphi2}).
Panels (c) and (d) report the phase jump $\delta\phi$ as a function of the rotation frequency $\omega$, for different barrier number (see legend). Panels (c) reports the result in the hydrodynamic regime, while panel (f) reports the Josephson regime data. Vertical dashed lines highlight the values of $\delta\phi_c$ for different $n$, while dashed horizontal line reports the value of $\delta\phi_c(1)$.
(e) Critical frequency $\omega_c$ as a function of the number of barriers $n$. Red (Blue) data reports the Josephson (hydrodynamic) results.
Full lines are guides to the eye. Circles show numerical results, while solid lines report the prediction of Eq.~(\ref{critfreq2}). Blue triangles report Eq.~(\ref{critfreq2}) for fixed values of $I_c(n)$ and $\delta\phi_c$ (to the $n = 1$ value). Red triangles reports the prediction of Eq.~(\ref{critfreq3}), valid in the Josephson regime. Parameters are $g = 10^4$, $V_0/\mu\sim1.8$, $\sigma/\xi_b\sim0.9$ in the Josephson regime and $g = 10^5$, $V_0/\mu\sim0.4$ and $\sigma/\xi_b\sim28$ in the hydrodynamic regime.
}
\label{Fig7}
\end{figure}
 The superfluid current $J$ is related to the phase drop $\delta\phi$ across each barrier by the current-phase relation. 
In the Josephson regime ($V_0/\mu \gtrsim 1$), it is $\delta\phi = \arcsin{(J/J_c)} -2\pi lJ$, where $J_c$ is the Josephson critical current (i.e. the maximum current which can be transported in the system) while $l$ is an adimensional kinetic inductance.
 For low $l $, the current thus reaches its maximum value $J_{\text{max}} \approx J_c$ at $\delta\phi = \pi/2$. 
 However, $\delta\phi$ can assume values larger than $\pi/2$, even in the Josephson regime. 
 In the case of a rotating ring, there is another constraint in the system: the maximum angular frequency $\omega_c$ at which the dynamical instability occurs. 
 We recall that $\delta \phi_c$ is the value of the phase drop observed when $\omega = \omega_c$.
 We thus have that, if $\delta \phi_c$ lies in the regime where the current {\it increases} with $\delta \phi$ (this corresponds to $\delta \phi \leq \pi/2$ in the Josephson regime), namely $d J/d \delta\phi \geq 0$, then reaching the maximum frequency corresponds to reaching the maximum current. 
 This is the case shown by the blue line in Fig.~\ref{Fig7}(a).
Instead, if $\delta \phi_c$ occurs in the regime where the current {\it decreases} with $\delta \phi$ (this corresponds to $\delta \phi > \pi/2$ in the Josephson regime, see also~\cite{Baratoff1970,Piazza2010}), namely $d J/d \delta\phi < 0$, then the maximum current is not observed at the maximum frequency.
The maximum current is reached at $\omega_{\rm max} < \omega_c$.
This is the case shown by the black and red lines in Fig.~\ref{Fig7}(a).
\subsection{Phase jump and instability}\label{AppPhase}

As discussed in Sec.~(\ref{Increasing}), the increase of the critical frequency $\omega_c(n)$ with $n$ is associated with the redistribution of the phase drop $\delta\phi(\omega,n)$ across each potential barrier.
Here, we provide a numerical analysis of Eqs.~(\ref{deltaphi}) and~(\ref{deltaphi2}), both in the hydrodynamic and Josephson regimes, and show that the critical value $\delta\phi_c=\delta\phi(\omega_c)$ depends only weakly on $n$.
Figure~(\ref{Fig7})(b) reports the prediction of Eq.~(\ref{deltaphi2}) for two different transport regimes: hydrodynamic ($V_0/\mu \ll 1$, $\sigma \gg \xi_b$, $\tilde{\sigma}\sim3\sigma$, blue line) and Josephson ($V_0/\mu > 1$, $\sigma \sim \xi_b$, $\tilde{\sigma}\sim6\,\xi_b$, red line). Dots represent numerical data, given by Eq.~(\ref{deltaphi}).
The phase jump $\delta\phi$ is observed to decrease for increasing $n$ and fixed $\Omega$, in agreement with Eq.~(\ref{deltaphi}). 
Figure~(\ref{Fig7})(c) and~(\ref{Fig7})(d) report the values of $\delta\phi(\Omega,n)$ as a function of $\Omega$ for the regimes of Fig.~(\ref{Fig7})(b) and for $n = 1,2,3$. The horizontal dashed line reports the value of $\delta\phi_c(1)$. Vertical colored lines mark the values of $\omega_c$ for $n=1,2,3$.
We find that $\delta\phi(\Omega,n)$ increases with $\Omega$ at fixed $n$.
Moreover, the critical value $\delta\phi_c = \delta\phi(\omega_c)$ is approximately constant with $n$, increasing only weakly. 
This supports the conclusion of Sec.~(\ref{Increasing}): increasing $n$ enhances stability because it reduces the phase jump per barrier, $\delta\phi$. If $\delta\phi_c$ decreased with $n$ as well, this argument would no longer hold, since the system would not be pushed further from the phase-slip threshold.
Figure~(\ref{Fig7})(e) reports the critical frequency $\omega_c$ as a function of $n$, for both the hydrodynamic (blue) and Josephson (red) regimes. Solid lines are Eq.~(\ref{critfreq2}) with $\tilde{\sigma} \sim 3\sigma$ (hydrodynamic) and $\tilde{\sigma}\sim6\xi_b$ (Josephson). Dots are numerical results. Blue triangles are Eq.~(\ref{critfreq2}) with $\delta\phi_c(n) = \delta\phi_c(1)$ and $I_c(n) = I_c(1)$. Red triangles are Eq.~(\ref{critfreq3}), obtained for the Josephson regime.
As we pointed out in Sec.~(\ref{Increasing}), keeping $\delta\phi_c$ and $I_c$ fixed to their $n = 1$ value still predicts a linear behavior. Inclusion of the full dependence results in the solid blue line, which is fully compatible with numerical data. 
On the other hand, the prediction of Eq.~(\ref{critfreq3}) is approximately consistent with the numerical result, giving the correct order of magnitude but the incorrect slope. 
This discrepancy arises because $\delta\phi_c$ increases slightly with $n$ at the critical point; as a result, Eq.~(\ref{critfreq3}) predicts an incorrect slope and appears as a secant line, crossing the numerical curve

\begin{figure}[t!]
\centering
\includegraphics[width=\columnwidth]{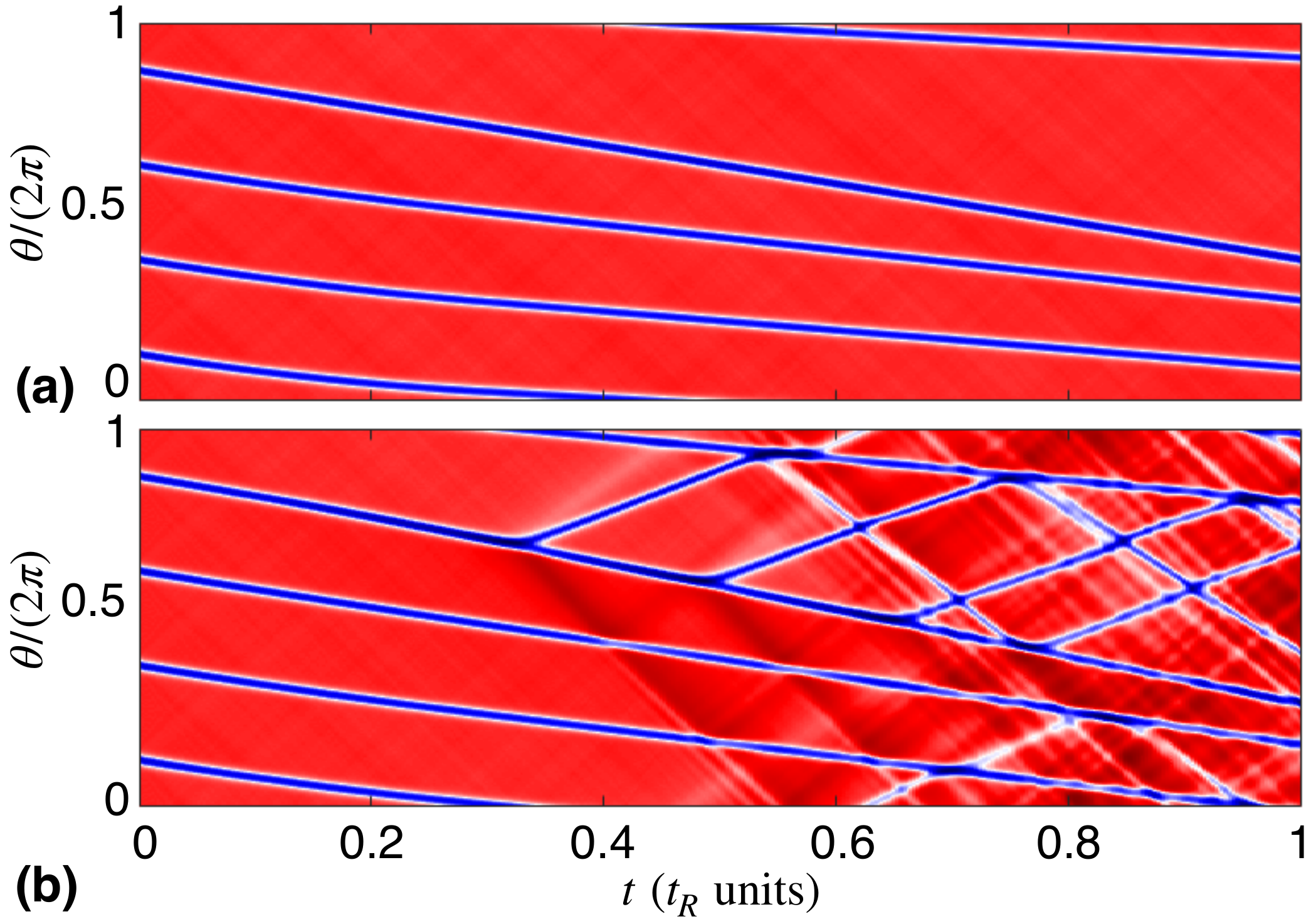}
\caption{Dynamical evolution of density profiles 
in the case when all barriers rotate at different angular velocities.  
In panel (a) all speeds are subcritical.
In panel (b) one barrier is overcritical. 
In both cases, see text for details on the speed values. 
Colorbar is the same as Fig.~\ref{Fig4}. 
The simulations of Fig.~\ref{CorrVEL} are stopped before the overlap of nearby barriers.
Parameters are set as $g = 1250$, $V_0/\mu \sim 0.9$, $\sigma/\xi \sim 1.2$.
}
\label{CorrVEL}
\end{figure}
 
 \subsection{Non-uniform barrier velocities}\label{app3}
We consider the case when each barrier rotates at a different angular velocity $\Omega_j$. 
Results are shown in Fig.~\ref{CorrVEL}.
If the angular rotation speed is subcritical for all barriers, then no soliton is observed during the time evolution: see for instance Fig.~\ref{CorrVEL}(a) where the velocities are 
$\omega_c/3, \omega_c/2, 2\omega_c/3$ and $\omega_c$, from bottom to top.
The unstable dynamics is qualitatively different from the case when all barriers rotate at the same velocity. 
In panel (b) we report the case of a dynamical quench of the above frequencies as $\omega\to\omega_c + \Delta\omega$ (with $\delta \omega\sim 0.2 \omega_c$), which causes only the top barrier to be driven at $\Omega_1 >\Omega_c = \nu-\omega_c$.
We observe that solitons are emitted only by this barrier, which acts as the strongest perturbation. 
This behavior is analogous to the case of non-uniform barrier heights or widths discussed in Sec.~IV\,A, where the instability was triggered by the "strongest'' barrier (largest $V_0$ or $\sigma$). 
\begin{figure}
\centering
\includegraphics[width=\columnwidth]{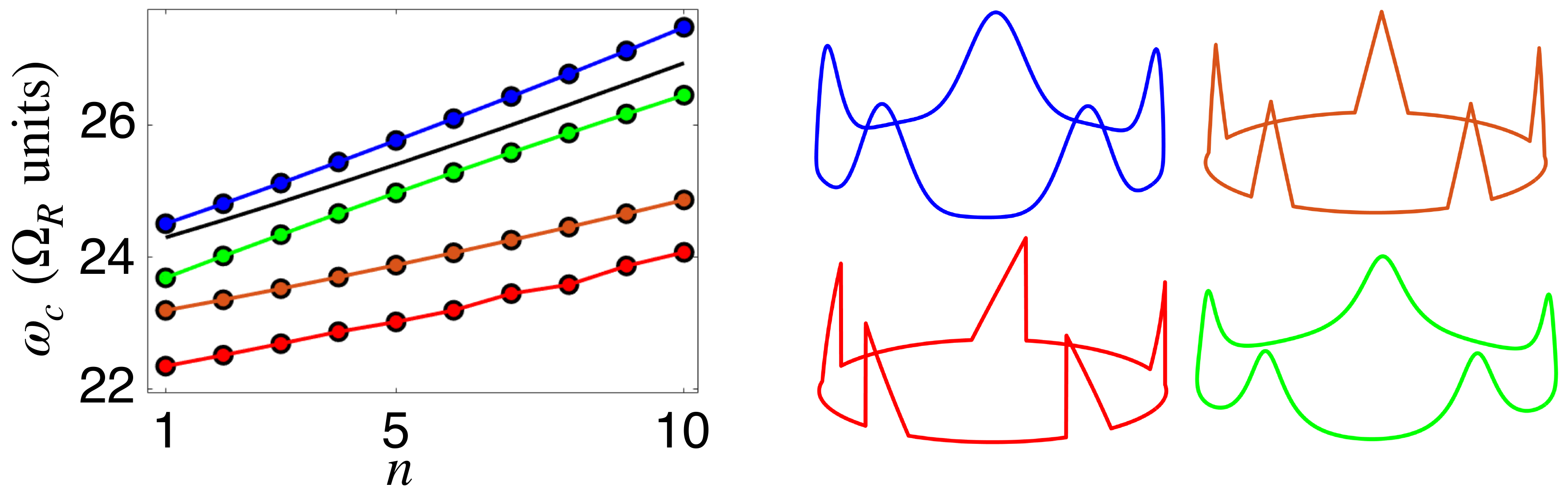}
\caption{The left panel reports critical frequency $\omega_c$ as a function of $n$ for barriers having different shapes: Gaussian (blue), triangular (orange), tilted triangular (red), Lorentzian (green)  
The corresponding potential $V(\theta)$ is shown on the right side, for the case $n=5$.
Lines crossing the points are guides to the eye.
The solid black line is the case of square barriers.
Data for different colors are shifted vertically for better visibility. 
Parameters are $g = 5000, V_0/\mu\sim 0.25$, $\sigma/\xi_b\sim6$, where $\sigma$ is the effective barrier half-width (Gaussian standard deviation, Lorentzian half-width, and half the triangular base).
}
\label{Fig8}
\end{figure}

\subsection{Independence of the stabilization effect from the barrier shape}\label{shapes}

Throughout this manuscript we have focused on square barriers in order to exploit analytical solutions of Eq.~(\ref{1DGPERes}).
In this section we show that our main conclusions are qualitatively insensitive to the barrier shape.
Figure~\ref{Fig8} shows the critical frequency $\omega_c$ calculated for different barrier shapes: Gaussian (blue), Lorentzian (green), triangular (orange), "grid-like"(red). Black solid line is the critical frequency calculated for a square-barrier potential as Fig.~\ref{Fig1}(a).    
Despite the different barrier shapes, we observe the same qualitative behavior as in the main paper: a linear increase with $n$ as long as nearby barriers are well-separated (i.e., $\theta_i -\theta_{i-1}>2 \xi_b,~\sigma$). 
This supports our claim that the stability of the system increases due to a redistribution of the phase jump $\delta\phi$ across the barriers when increasing $n$.


\begin{thebibliography}{100}
\bibitem{footnotecoeff}
the coefficients entering Eqs.~(\ref{solutionsins}) and~(\ref{solutionsout}) are:
\begin{eqnarray}
&& \Delta = \left(\rho_{0}-\frac{2(\mu_{0}-V_{0})}{g}\right)^{2} - \frac{4J^{2}}{g\,\rho_{0}} \nonumber \\  
&& \rho_{1} = \tfrac{3}{2}\rho_{0} - \frac{\mu_{0}-V_{0}}{g} - \tfrac{\sqrt{\Delta}}{2} \nonumber \\
&& \alpha_{1}^{2} = g(\sqrt{\Delta}+\rho_{1}) \nonumber \\
&& m_{1} = \dfrac{\sqrt{\Delta}}{\sqrt{\Delta}+\rho_{1}} \nonumber \\
&& \rho_{2} = \sqrt{2}\,\!\bigl(\rho_{0}^{2} - \rho_{0}\tfrac{\mu_{0}-V_{0}}{g} + \tfrac{J^{2}}{2g\,\rho_{0}}\bigr)^{1/2} \nonumber \\
&& \alpha_{2}^{2} = 4g\rho_{2} \nonumber \\
&& m_{2} = \dfrac{\rho_{2}-\rho_{1}-\tfrac{\sqrt{\Delta}}{2}}{2\rho_{2}}  \nonumber \\
&& c_{b} = \rho_{b} - \dfrac{2\,\xi_{b}^2J^2}{{g}\rho_{b}} \nonumber \\
&& \phi_{0}=\operatorname{arctanh}(\mathsmaller{\sqrt{(\rho_{\sigma}-2\xi_{b}^{2}J^{2}/\rho_{b})/c_{b}}}) \nonumber \\
&& \rho_{\sigma} = \dfrac{g}{2V_{0}}\!\left(\rho_{0}^{2}-\rho_{b}^{2}
+ \dfrac{2\mu_{0}\rho_{b}}{g}
- \dfrac{(\mu_{0}-V_{0})\rho_{0}}{g}
+ \dfrac{J^{2}}{g}\left(\rho_{b}^{-1}-\rho_{0}^{-1}\right)\right). \nonumber
\end{eqnarray}
\bibitem{note}
In all cases, $\tilde{\sigma}$ is chosen sufficiently large to cover the region where the density is depleted, i.e., $\tilde{\sigma} \sim 3$--$10\,\xi_b$.
\bibitem{notedisorder}
The speckle is generated on the discrete grid
$\theta_j = 2\pi j/N$ by Fourier filtering a complex random field with a
top–hat aperture in momentum space. Its intensity
$I(\theta_j)=|\psi(\theta_j)|^2$ is shifted to be non–negative and
re-normalized such that its maximum value is unity.
\end{thebibliography}
\end{document}